\documentclass[journal,comsoc]{IEEEtran}

\usepackage{amssymb,color}
\usepackage{amsmath}
\usepackage{bbm}
\usepackage{graphicx}
\usepackage{epstopdf}
\usepackage{amsthm,amsfonts}
\usepackage{subcaption}

\usepackage{accents}

\usepackage{algorithm}
\usepackage{algorithmic}
\usepackage{psfrag}
\usepackage{cite}
\usepackage{pifont}
\usepackage[font={small}]{caption}

\allowdisplaybreaks

\makeatletter
\newcommand{\oset}[3][0ex]{%
  \mathrel{\mathop{#3}\limits^{
    \vbox to#1{\kern-2\ex@
    \hbox{$\scriptstyle#2$}\vss}}}}
\makeatother

\begin{document}
\title{Meta Distribution of Downlink Non-Orthogonal Multiple Access (NOMA) in Poisson Networks} 

\author{
\IEEEauthorblockN{\large  Konpal Shaukat Ali$^{\ast}$, Hesham ElSawy$^{\dagger}$, and Mohamed-Slim Alouini$^{\ast}$}

\thanks{$^{\ast}$ The authors are with the Computer, Electrical, and Mathematical Sciences and Engineering (CEMSE) Divison, King Abdullah University of Science and Technology (KAUST), Thuwal, Makkah Province, Saudi Arabia. (Email: \{konpal.ali, slim.alouini\}@kaust.edu.sa)

$^{\dagger}$ The author is with the Department of Electrical Engineering, King Fahd University of Petroleum and Minerals (KFUPM), Dhahran, Saudi Arabia. (Email: hesham.elsawy@kfupm.edu.sa)

 }}

\maketitle

\begin{abstract} 
We study the meta distribution (MD) of the coverage probability (CP) in downlink non-orthogonal-multiple-access (NOMA) networks. Two schemes are assessed based on the location of the NOMA users: 1) anywhere in the network, 2) cell-center users only. The moments of the MD for both schemes are derived and the MD is approximated via the beta distribution. Closed-form moments are derived for the first scheme; for the second scheme exact and approximate moments, to simplify the integral calculation, are derived. We show that restricting NOMA to cell-center users provides significantly higher mean, lower variance and better percentile performance for the CP.

\end{abstract}


\section{Introduction}

Conventionally, orthogonal multiple access (OMA) is used for transmissions to different users (UEs) served by the same base station (BS). OMA assigns different time-frequency resource blocks (TF-RBs) to each UE to avoid intracell interference. However, spectrum scarcity and the increasing capacity demand call for more efficient spectrum utilization. In this regard, {non-orthogonal multiple access (NOMA) is a technique that improves spectral efficiency by superposing the messages of multiple UEs on one TF-RB. Successive interference cancellation (SIC) is used for NOMA decoding.} {The superiority of NOMA over OMA schemes {in a noise-limited regime} is well established from an information theoretic perspective~\cite{book_fundWirelessComm}.}





{Using stochastic geometry, the superiority of NOMA has also been established for large-scale interference prone networks}~\cite{myNOMA_tcom,myNOMA_icc,N6,my_nomaMag}. Such studies usually focus on the spatially averaged coverage probability (SCP), which averages the coverage probability (CP) over all fading, activity, and network realizations. However, network operators are usually more interested in the percentile performance of UEs, where the fading and activity change while the network realization is kept constant. {The CP given a fixed network realization is defined as the conditional CP (CCP) \cite{meta_mh_bipolarAndCell}}. The {complementary cdf} of the {CCP}, denoted as the meta distribution (MD), reveals the percentile performance across an arbitrary network realization. \cite{metaNOMA_EH} studies the MD for uplink and downlink NOMA with NOMA UEs located everywhere in the network; however, the joint decoding associated with SIC is not taken into account.  



This letter characterizes the MD in downlink cellular networks for two NOMA schemes, namely, everywhere NOMA (E-NOMA) and cell-center NOMA (C-NOMA). E-NOMA utilizes NOMA for UEs located everywhere in the network \cite{my_nomaMag,metaNOMA_EH}, while C-NOMA restricts NOMA to cell-center UEs only \cite{myNOMA_tcom,myNOMA_icc}. {We derive closed-form expressions for the moments of the MD in E-NOMA. Integral expressions are obtained for the moments in C-NOMA; consequently, we propose accurate approximate moments to simplify the integral calculation.} The MD is then approximated using the beta distribution via moment matching to characterize the UEs percentile performance. Different from~\cite{metaNOMA_EH} we {derive and compare the statistics of the MD} for two NOMA schemes, and consider joint decoding {for all SIC phases}. {To the best of our knowledge, NOMA works in the literature employ one scheme and do not compare different schemes.} Our results show that C-NOMA not only provides higher SCP, but also reduces the variance of the CP across the UEs in the network when compared to the E-NOMA. 


\vspace{-.08in}
\section{System Model}\label{SysMod}
\vspace{-.05in}
We consider a downlink cellular network where BSs are distributed according to a homogeneous PPP $\Phi$ with intensity $\lambda$. Each BS serves {$N$ UEs} in one TF-RB by multiplexing the signals for each UE with different power levels using a total power budget $P=1$. A Rayleigh fading environment is assumed such that the fading coefficients {are i.i.d.~with} a unit mean exponential distribution. A power-law path-loss model is considered where the signal decays at the rate $r^{-\eta}$ with distance $r$, $\eta>2$ denotes the path-loss exponent and $\delta=\frac{2}{\eta}$.




SIC requires ordering the UEs according to some measure of link strength \cite{myNOMA_tcom}. For $i \in \{1, \ldots ,N\}$, the $i^{th}$ strongest UE is referred to as $\text{UE}_i$. In this work, we order the UEs based on the link distance $R$. The ordered link distance of $\text{UE}_i$ is denoted by $R_i$; consequently, $\text{UE}_i$ is nearer to the BS {and therefore stronger} than $\text{UE}_{j}$ for $i<j$ {(i.e., $R_i<R_j$)}. Exploiting SIC, $\text{UE}_i$ decodes and cancels messages intended for all weaker UEs before decoding its own message. On the other hand, messages for stronger UEs are treated as noise and contribute to the intracell interference. We incorporate imperfect SIC into our analysis by considering a fraction $\beta$ of residual intracell interference from the canceled messages of weaker UEs. Let $P_i$ and $\log(1+\theta_i)$ denote the power allocated and target rate for  $\text{UE}_i$; {the corresponding signal-to-interference ratio (SIR) threshold for the message of $\text{UE}_i$ is $\theta_i$. Note that due to the power budget, $\sum_{i=1}^N P_i=1$.} For feasible SIC, proper {resource allocation (RA), i.e.,} power allocation and rate adaptation (e.g., $P_i \leq P_j$ and/or $\theta_i \geq \theta_j$ for $i<j$), for all UEs is required. 

\textbf{\textbf{\emph{Lemma 1}:}} {For any ascending ordered statistic like $R_i$, based on the statistics of the unordered counterpart $R$, the pdf is} 
\begin{align}
f_{R_i}(r)=\binom{\!N-1\!}{\!i-1\!} N f_{R}(r)  \left( F_{R}(r) \right)^{i-1} \! \left( 1 \!-\!F_{R}(r) \right)^{N-i}. \label{f_Ri_gen}
\end{align}
In terms of {components larger than $i$}, \eqref{f_Ri_gen} can be rewritten as
\begin{align}
f_{R_i}(r)= f_{\widehat{R}_i}(r) + \sum_{m=i+1}^N \binom{m-1}{i-1} (-1)^{m-i} f_{\widehat{R}_m}(r), \label{ord_weak}
\end{align}
where $f_{\widehat{R}_j}(r)=\binom{N-1}{j-1} N f_R(r) (F_R)^{j-1}$ for $i \leq j \leq N$. In terms of {components smaller than $i$}, \eqref{f_Ri_gen} can be rewritten as
\begin{align}
f_{R_i}(r)= f_{\widetilde{R}_i}(r) + \sum_{m=1}^{i-1} \frac{(N-m)! (-1)^{i-m}}{(m-1)! (i-m)!}  f_{\widetilde{R}_m}(r), \label{ord_strong}
\end{align}
where $f_{\widetilde{R}_j}(r)=\binom{N-1}{j-1} N f_R(r) (1-F_R)^{N-j}$ for $1 \leq j \leq i$.

We denote the distance between a BS and its nearest neighboring BS by $\rho$. Since $\Phi$ is a PPP, the pdf of $\rho$ is $f_{\rho}(x)=2\pi \lambda x e^{-\pi \lambda x^2},$ $x\geq 0$.
Consider a disk around each BS located at $\textbf{x}$ with radius $\rho/2$, i.e., $b(\textbf{x},\rho/2)$; we refer to this as the in-disk. The in-disk is the largest disk centered at a BS that fits inside its Voronoi cell. We study and compare NOMA for the following two schemes. 
\subsubsection{\textbf{Everywhere NOMA (E-NOMA)}}
{$N$ UEs} are distributed uniformly and independently in each Voronoi cell. Consequently, the distribution of the unordered link distance $R$ follows $f_{R}(r)=2\pi\lambda r e^{-\pi\lambda r^2}$, $r\geq 0$. Using this pdf and its cdf $F_R(r)$, the ordered distance distribution $f_{R_i}(r)$, $r\geq 0$, in the E-NOMA scheme follows \eqref{f_Ri_gen}.

\subsubsection{\textbf{Cell-Center NOMA(C-NOMA)}}
{$N$ UEs} are distributed uniformly and independently in the in-disk $b(\textbf{x},\rho/2)$ of each BS at \textbf{x} \cite{myNOMA_icc}. Consequently, the link distance $R$, conditioned on $\rho$, follows $f_{R\mid \rho}(r \mid \rho)=\frac{8r}{\rho^2}$, $0 \leq r \leq \frac{\rho}{2}$. 
Using \eqref{f_Ri_gen} the pdf of $R_i$, conditioned on $\rho$, in the C-NOMA scheme follows
{\begin{align}
{f_{R_i \mid \rho}(r \! \mid \! \rho)\!\! =\!\! \binom{\!N\!-\!1\!}{\!i\!-\!1\!}\!\! \frac{8rN }{\rho^2} \!\! \left(\! \frac{4r^2}{\rho^2} \!\right)^{\!i\!-\!1\!} \!\! \!\left(\! 1\! \!-\! \! \frac{4r^2}{\rho^2}\! \right)^{\!N-i}\!\!\!, 0 \! \leq \! r \! \leq \! \frac{\rho}{2}}. \label{f_Ri}
\end{align}}

\textbf{\emph{Remark:}} C-NOMA restricts the link distance to $\rho/2$; the notion is that NOMA is better suited for UEs that are closer to the serving BS. UEs with relatively larger link distances are better served in their own resource block without sharing~\cite{myNOMA_tcom}.

\section{SIR Analysis}

{SIC requires a UE to successfully decode all of the messages intended for weaker UEs.} Consider a randomly selected BS located at $\textbf{x}_0$ and its associated UEs; the SIR at $\text{UE}_i$ of the message intended for $\text{UE}_j$ for {$i\leq j \leq N$} is
{\begin{align*}
{\rm SIR}_j^i=\!\!\frac{h_i R_i^{-\eta} P_j}{ {h_i R_i^{-\eta}  \Bigg( \sum\limits_{m=1}^{j-1} P_m \!\!+ \beta \!\! \sum\limits_{k=j+1}^{N} \!\! P_k \Bigg)} \!\! +  \!\! {\sum\limits_{\textbf{x} \in {\Phi} \backslash \textbf{x}_0}  g_{\textbf{y}_i} {\|\textbf{y}_i\|}^{-\eta}} },
\end{align*}}
\noindent where $\textbf{y}_i=\textbf{x}-\textbf{u}_i$, $\textbf{u}_i$ is the location of $\text{UE}_i$, $\|\cdot\|$ denotes the Euclidean norm, and $h_i$ ($g_{\textbf{y}_i}$) is the fading power gain from the serving (interfering) BS to $\text{UE}_i$.

{Accordingly, due to SIC decoding, coverage at $\text{UE}_i$ is defined via the following joint event}
{\begin{align}
&C_i \!\!=\!\!  \bigcap\limits_{j=i}^N \left\lbrace {\rm SIR}_j^i\!>\! \theta_j \right\rbrace \!\!=\!\! \bigcap\limits_{j=i}^N \left\lbrace \! h_i  \!>\!  R_i^{\eta} \frac{\theta_j}{\tilde{P}_j} \!\! \sum\limits_{\textbf{x} \in {\Phi}\backslash \textbf{x}_0}  \!\! g_{\textbf{y}_i} {\|\textbf{y}_i\|}^{-\eta} \! \right\rbrace \! , \label{C_i_new}
\end{align}}
{$\text{where }\tilde{P}_j= P_j - \theta_j \! \left(\! \sum\limits_{m=1}^{j-1} P_m \!+\! \beta \!\! \sum\limits_{k=j+1}^{N} \!\! P_k \! \right)\!$.} We rewrite \eqref{C_i_new} as $C_i =  \left\lbrace h_i> R_i^{\eta} M_i \sum\limits_{\textbf{x} \in {\Phi}}  g_{\textbf{y}_i} {\|\textbf{y}_i\|}^{-\eta}   \right\rbrace$ using {$M_i$$=$$\max\limits_{i \leq j \leq N} \frac{\theta_j }{\tilde{P}_j }$}.


For a fixed, yet arbitrary, realization of the network, the CCP of $\text{UE}_i$ {in a randomly selected cell, ${\mathcal{P}_{C_i}}$, is}
\begin{align}
{\mathcal{P}_{C_i}=}\mathbb{P}(C_i | \Phi) &\stackrel{(a)}=\mathbb{E}_{g_{\textbf{y}_i}} \Bigg[ \exp  \Bigg( -R_i^{\eta} M_i \sum\limits_{\textbf{x} \in {\Phi} \backslash \textbf{x}_0 }  g_{\textbf{y}_i} {\|\textbf{y}_i\|}^{-\eta} \Bigg)  \mid \Phi \Bigg] \nonumber\\
& \stackrel{(b)}= \prod\limits_{\textbf{x} \in {\Phi} \backslash \textbf{x}_0} \frac{1}{1+R_i^{\eta} M_i {\|\textbf{y}_i\|}^{-\eta}},
\end{align}
where $(a)$ follows using the cdf of $h_i\sim \exp(1)$ and $(b)$ follows from the MGF of the independent RVs $g_{\textbf{y}_i} \sim \exp(1)$. 

Denote the $b^{th}$ moment of the CCP of $\text{UE}_i$ across all links in {an arbitrary fixed realization} of the network by $\mathcal{M}_{i,b}$. Then,
\begin{align}
&\mathcal{M}_{i,b}= \mathbb{E} \Bigg[  \prod\limits_{\textbf{x} \in {\Phi} \backslash \textbf{x}_0}  {\left(1+R_i^{\eta} M_i {\|\textbf{y}_i\|}^{-\eta} \right)}^{-b}      \Bigg]. \label{M_ib}
\end{align} 
\textbf{\emph{Remark:}} If $\tilde{P}_j\!<\!0$, the CCP is zero. Henceforth we assume RA such that $\tilde{P}_j \geq0$.\\ 
{\textbf{\emph{Note:}} If $b=1$ in \eqref{M_ib}, we obtain the SCP of $\text{UE}_i$.}

Through moment matching, the MD of $\text{UE}_i$ is approximated using the beta distribution~\cite{meta_mh_bipolarAndCell} as follows
\begin{align}
{\bar{F}_{{\mathcal{P}_{C_i}}}(\alpha)} {=\mathbb{P}\left(\mathcal{P}_{C_i} > \alpha \right)} \approx 1 - \mathcal{I}_{\alpha}\left( \frac{\beta_i \mathcal{M}_{i,1}}{1-\mathcal{M}_{i,1}}, \beta_i \right), \label{meta_beta}
\end{align}
where $\beta_i$=$\frac{(\mathcal{M}_{i,1}-\mathcal{M}_{i,2})(1-\mathcal{M}_{i,1})}{\mathcal{M}_{i,2}-\mathcal{M}_{i,1}^2}$ and $\mathcal{I}_{\alpha}(a,b)=\int_0^{\alpha} l^{a-1} (1-l)^{b-1} dl$. 
The variance {of the MD of $\text{UE}_i$} is defined as
\begin{align}
\sigma_i^2={\mathcal{M}_{i,2}-\mathcal{M}_{i,1}^2}.
\end{align}

The ordered relative distance process (RDP) for $\text{UE}_i$, which is the RDP in \cite{mh_Asymptotics} using ordered link distance, is 
\begin{align}
\mathcal{R}_i=\{{x \in \Phi \backslash \{x_0\}}: R_i/\|\textbf{y}_i\|\}.
\end{align}
Using the PGFL of the PPP in $(a)$, the PGFL of $\mathcal{R}_i$ is
\begin{align}
&\mathcal{G}_{\mathcal{R}_i}[f] \stackrel{\triangle}= \mathbb{E} \Bigg[\prod\limits_{x \in \mathcal{R}_i} f(x) \Bigg] \nonumber = \mathbb{E} \Bigg[\prod\limits_{{x \in \Phi \backslash \{x_0\}}} f\left(\frac{R_i}{\|\textbf{y}_i\|}\right) \Bigg] \nonumber \\
&\stackrel{(a)}=\mathbb{E}_{R_i} \left[\exp\left(-2 \pi \lambda \int_{R_i}^\infty \left(1-f\left(\frac{R_i}{a} \right) \right) a \; da \right) \right].\label{pgfl_rdp_gen}
\end{align}

Using the ordered RDP for $\text{UE}_i$, the expectation in \eqref{M_ib} can also be evaluated as 
\vspace{-.03in}
\begin{align}
&\mathcal{M}_{i,b}= \mathbb{E} \Bigg[  \prod\limits_{\textbf{y} \in {\mathcal{R}_i}}  {\left(1+ M_i y^{\eta} \right)^{-b}}      \Bigg] . \label{M_ib_RDP}
\end{align}
\subsubsection*{1) E-NOMA Scheme}
{We characterize the PGFL of the ordered RDPs and obtain closed for expressions for $\mathcal{M}_{i,b}$.}

{\textbf{\emph{Lemma 2:}} The PGFL of $\mathcal{R}_i$ for $1\leq i \leq N$ in E-NOMA is
\begin{align}
&\mathcal{G}_{\mathcal{R}_i}[f]= \mathcal{G}_{\mathcal{\widetilde{R}}_i}[f] + \sum_{m=1}^{i-1} \frac{(N-m)! (-1)^{i-m}}{(m-1)! (i-m)!}  \mathcal{G}_{\widetilde{R}_m}[f] , \label{G_ri}\\
&\text{where for } 1 
\leq j \leq i \nonumber \\ 
& \mathcal{G}_{\widetilde{\mathcal{R}}_j}[f]=  \frac{ \binom{N-1}{j-1} N}{(N-j+1) + 2\int_1^\infty \left(1-f\left({y^{-1}} \right)\right)  y \; dy }. \label{G_rTilde_j}
\end{align}
\emph{Proof:} We obtain \eqref{G_ri} using \eqref{ord_strong} in \eqref{pgfl_rdp_gen}. Also using \eqref{pgfl_rdp_gen},
\begin{align*}
&\mathcal{G}_{\widetilde{\mathcal{R}}_j}[f]= \int_0^{\infty} \!\! f_{\widetilde{R}_j}(x) \exp\left(\!-2 \pi \lambda \int_{R_i}^\infty \left(1-f\left(\frac{x}{a} \right) \! \right) a \; da \! \right) \! dx \nonumber \\
&\stackrel{(a)}= {\binom{\!N\!-\!1\!}{\!j\!-\!1\!}\! \pi\lambda N \int_0^{\infty} \!\! e^{-2\pi \lambda m \! \int\limits_1^\infty \! \left(\!1 \! - \! f\left({y^{-1}} \right) \! \right)  y \; dy \!  } e^{-\pi\lambda (N-j+1) m} dm }
\end{align*}
 where $(a)$ is obtained by changing variables and \eqref{G_rTilde_j} is obtained using the MGF of $m\sim \exp(\pi\lambda (N-j+1))$. \qed
}

{\textbf{\emph{Corollary 1:}} $\mathcal{M}_{i,b}$ for $1 \leq i \leq N$ in E-NOMA is
\begin{align}
&\mathcal{M}_{i,b}= \widetilde{\mathcal{M}}_{i,b} + \sum_{m=1}^{i-1} \frac{(N-m)! (-1)^{i-m}}{(m-1)! (i-m)!} \widetilde{\mathcal{M}}_{m,b} , \label{M_ib_vCell} \\
&\text{where for } 1 \leq j \leq i \nonumber \\
& \widetilde{\mathcal{M}}_{j,b}= \binom{N-1}{j-1} \frac{N}{N-j+ {}_2F_1\left(b,-\delta,1-\delta,-M_i \right) } \label{M_ibTilde_vCell}.
\end{align}
\emph{Proof:} \eqref{M_ib_vCell} is obtained using \eqref{G_ri}, where we define using \eqref{M_ib_RDP}
\begin{align*}
&\widetilde{\mathcal{M}}_{j,b} \!=\! \mathcal{G}_{\widetilde{\mathcal{R}}_j}\left[\frac{1}{\!{(1\!+\!M_i y^{\eta})}^{b}\!} \right] \!\stackrel{(a)}=\!  \frac{ \binom{N-1}{j-1} N}{N\!-\!j\!+\!1 \!+\! 2 \! \int\limits_1^\infty \!\left(\!1 \!-\!\left(\!1\!+\!M_i y^{-\eta} \!\right)^{-b} \!\right)\!  y  dy }.
\end{align*}
We obtain $(a)$ using \eqref{G_rTilde_j}, and \eqref{M_ibTilde_vCell} follows by $y\rightarrow g^{-1}$.  \qed}

%

\subsubsection*{2) C-NOMA Scheme} {We obtain integral expressions for $\mathcal{M}_{i,b}$. We also propose approximate PGFLs of the ordered RDP and use these to evaluate $\mathcal{M}_{i,b}$ in a simpler form.  }

\textbf{\emph{Lemma 3:}} The $b^{th}$ moment of the CCP for $\text{UE}_i$ in the C-NOMA scheme is 
\begin{align}
&\!\mathcal{M}_{i,b} \!\approx \!  \mathbb{E}_{\rho,R_i} \!\Bigg[ \! {  e^{  -\!2\pi \! \lambda \!\!\!\! \int\limits_{\rho-R_i}^{\infty} \!\! \!\left(\!\! 1 \!-\! {\left(1 \! + \!  \frac{M_i R_i^{\eta}}{r^{\eta}} \right)}^{\!-b}\! \right) \! r dr  \!\!}}     {\left(\!1\!\!+\!  \frac{M_i  R_i^{\eta}}{\rho^{\eta}} \! \right)}^{\!-b} \! \Bigg] \!. \label{M_i_CC}
\end{align}
\textbf{\emph{Proof:}} In the C-NOMA model each UE is conditioned to have an interferer $\rho$ away from the serving BS. Hence, using \eqref{M_ib}
\begin{align*}
&\mathcal{M}_{i,b} \!=\!\mathbb{E} \Bigg[\! \!\!  \prod\limits_{\substack{\textbf{x}\in\Phi \backslash \textbf{x}_0 \\ \|\!\textbf{x} -\textbf{x}_0 \! \|>\rho}} \!\!\!\!  {\left(\! 1+ M_i\frac{R_i^{\eta}}{{\|\textbf{y}_i\|}^{\eta}} \! \right)}^{-b}  \!\! \!\!\!\! \prod\limits_{\substack{\textbf{x}\in\Phi \backslash \textbf{x}_0 \\ \|\! \textbf{x} -\textbf{x}_0\!\|=\rho}} \! \!\! \!  {\left(\!1+ M_i\frac{R_i^{\eta}}{{\|\textbf{y}_i\|}^{\eta}}  \!\right)}^{-b}        \Bigg] .
\end{align*}
We obtain the first term in \eqref{M_i_CC} using the PGFL of the PPP and the guard zone $b(\textbf{u}_i,\rho-R_i)$ in the C-NOMA scheme. The average location of a UE distributed uniformly in the in-disk is the center of the disk, i.e, $\textbf{x}_0$. Accordingly, we approximate the average distance between a UE and the BS $\rho$ away from $\textbf{x}_0$ as $\rho$; hence,  the second term in \eqref{M_i_CC} is obtained. This approximation has been validated to be tight in  \cite{myNOMA_tcom,myNOMA_icc}. \qed

Consider the following two approximations:\\
$\bullet$ \textbf{A1:} $\text{UE}_i$ is guaranteed to have no interfering BS in $b(\textbf{u}_i,R_i)$, which is not the largest guard zone around the UE.\\
$\bullet$ \textbf{A2:} Deconditioning on the BS $\rho$ away from the serving BS.

\textbf{\emph{Remark:}} The two approximations have opposing effects; \textbf{A1} overestimates intercell interference while \textbf{A2} underestimates it.

Calculating $\mathcal{M}_{i,b}$ using Lemma 3 requires a triple integral. {However, exploiting} \textbf{A1} and \textbf{A2}, we provide an approximation to calculate $\mathcal{M}_{i,b}$ that requires a single integration.

{ \textbf{\emph{Lemma 4:}} Using \textbf{A1} and \textbf{A2}, the PGFL of $\mathcal{R}_i$ conditioned on $\rho$ for $1 \leq i \leq N$ in the C-NOMA scheme is 
\begin{align}
&\mathcal{G}_{\mathcal{R}_i \mid \rho}[f]= \mathcal{G}_{\mathcal{\widehat{R}}_i \mid \rho}[f] \! + \! \! \sum_{m=i+1}^{N} \! \binom{\!m\!-\!1\!}{\!i\!-\!1\!} (-1)^{m-i} \mathcal{G}_{\widehat{R}_m \mid \rho }[f] , \!\! \label{G_ri_CC}\\
&\text{where for } i 
\leq j \leq N \nonumber \\ 
& \mathcal{G}_{\widehat{\mathcal{R}}_j \mid \rho}[f] \!= \!    \frac{ \binom{\!N\!-\!1\!}{\!j\!-\!1\!} \! \left( \! \Gamma(j) \! - \! \Gamma \left(j, \! \frac{\pi\lambda\rho^2}{2} \!  \! \int_1^\infty \! \! \left(\!1\!- \!f \!\left( \! \frac{1}{y} \! \right) \! \right) \! y dy \! \right) \! \right) }{ \frac{1}{N}  \left(\frac{\rho^2}{2} \pi\lambda \int_1^\infty \! \left(1 \! - \! f\left( \! \frac{1}{y} \! \right)\right) \! y dy \right)^j }. \label{G_rHat_j}
\end{align}
\emph{Proof:} We obtain \eqref{G_ri_CC} using \eqref{ord_weak} in \eqref{pgfl_rdp_gen}. Also using \eqref{pgfl_rdp_gen},
\begin{align*}
&\mathcal{G}_{\widehat{\mathcal{R}}_j \mid \rho }[f]= \int_0^{\infty} \!\! f_{\widehat{R}_j}(x) \exp\left(\!-2 \pi \lambda \int_{R_i}^\infty \left(1-f\left(\frac{x}{a} \right) \! \right) a \; da \! \right) \! dx \nonumber \\
&\stackrel{(a)}={\binom{\!N\!-\!1\!}{\!j\!-\!1\!} N \frac{ 4^j }{\rho^{2j}}  \int_0^{\frac{\rho^2}{4}} \!\! e^{-2\pi \lambda m  \int\limits_1^\infty  \left(\!1 \! - \! f\left({y^{-1}} \right) \! \right)  y \; dy } m^{j-1} dm }
\end{align*}
$(a)$ follows by changing variables, and \eqref{G_rHat_j} by integration. \qed 
}\\
{We approximate $\mathcal{M}_{i,b}$ by substituting the approximate PGFL of $\mathcal{R}_i$, conditioned on $\rho$, into \eqref{M_ib_RDP} and averaging over $\rho$.}

{\textbf{\emph{Corollary 2:}} Using \textbf{A1} and \textbf{A2}, $\mathcal{M}_{i,b}$ for $1 \leq i \leq N$ in C-NOMA is
\begin{align}
&\mathcal{M}_{i,b}= \widehat{\mathcal{M}}_{i,b} + \sum_{m=i+1}^{N} \binom{m-1}{i-1} (-1)^{m-i}   \widehat{\mathcal{M}}_{m,b} ,\label{M_ib_CC} \\
&\text{where for } i \leq j \leq N \nonumber \\
& \widehat{\mathcal{M}}_{j,b} \!=\! \mathbb{E}_{\rho} \!\! \left[ \! \frac{ \! \Gamma(j) \!\!- \!\! \Gamma \! \left(j, \! \! \frac{\pi\lambda  \rho^2}{4} \!  \! \left({}_2F_1 \!\left(b,\!-\delta,1\!-\!\delta,\!-M_i \right)\!- \!1 \! \right) \!\right) \! }{ \frac{(\pi \lambda)^j}{ \binom{N-1}{j-1} N } \frac{{ \rho^{2j}}}{4^j}   \left({}_2F_1\left(b,\!-\delta,1\!-\!\delta,\!-M_i \right)\!- \!1 \right)^j} \! \right] \! \!. \label{M_ibHat}
\end{align}
\emph{Proof:} {\eqref{M_ib_CC} is obtained using \eqref{G_ri_CC} where we define using \eqref{M_ib_RDP}
\begin{align*}
&\widehat{\mathcal{M}}_{j,b} = \mathbb{E}_{\rho} \left[ \mathcal{G}_{\widehat{\mathcal{R}}_j \mid \rho}[(1+M_i y^{\eta})^{-b}] \right] \\
&\stackrel{(a)}= \mathbb{E}_{\rho} \!\! \left[ \frac{ \binom{\!N\!-\!1\!}{\!j\!-\!1\!} \! \left( \! \Gamma(j) \! - \! \Gamma \left(j, \! \frac{\pi\lambda\rho^2}{2} \!  \! \int_1^\infty \! \! \left(\!1\!- {(1+M_i y^{-\eta})^{-b}}  \right) \! y dy \! \right) \! \right) }{ \frac{1}{N}  \left(\frac{\rho^2}{2} \pi\lambda \int_1^\infty \! \left(1 \! - {(1+M_i y^{-\eta})^{-b}}  \right) \! y dy \right)^j } \right].
\end{align*}
We obtain $(a)$ using \eqref{G_rHat_j}, and \eqref{M_ibHat} follows by $y\rightarrow g^{-1}$.  \qed}
}

\section{Results}
In this section, we select the following parameters: $\lambda=10$, $\eta=4$, $\beta=0$ and {$N=2$}, {unless stated otherwise}. Simulations are repeated {50,000} times. Since the power budget is $P=1$, $P_2=1-P_1$. Unless stated otherwise, Lemma 3 is used for the moments of the CCP in the C-NOMA model.


\begin{figure}[htb]
\centering\includegraphics[width=0.3\textwidth]{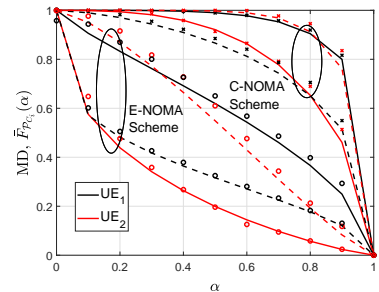}
\caption{MD vs. $\alpha$ with $\theta_1=1$ and $\theta_2=0.5$. Solid lines represent $P_1=0.5$, dashed $P_1=0.1$, markers show Monte Carlo simulations.}\label{meta}
\end{figure}

\begin{figure*}[t]
\begin{minipage}[t]{0.33\linewidth}
\centering\includegraphics[width=1\textwidth]{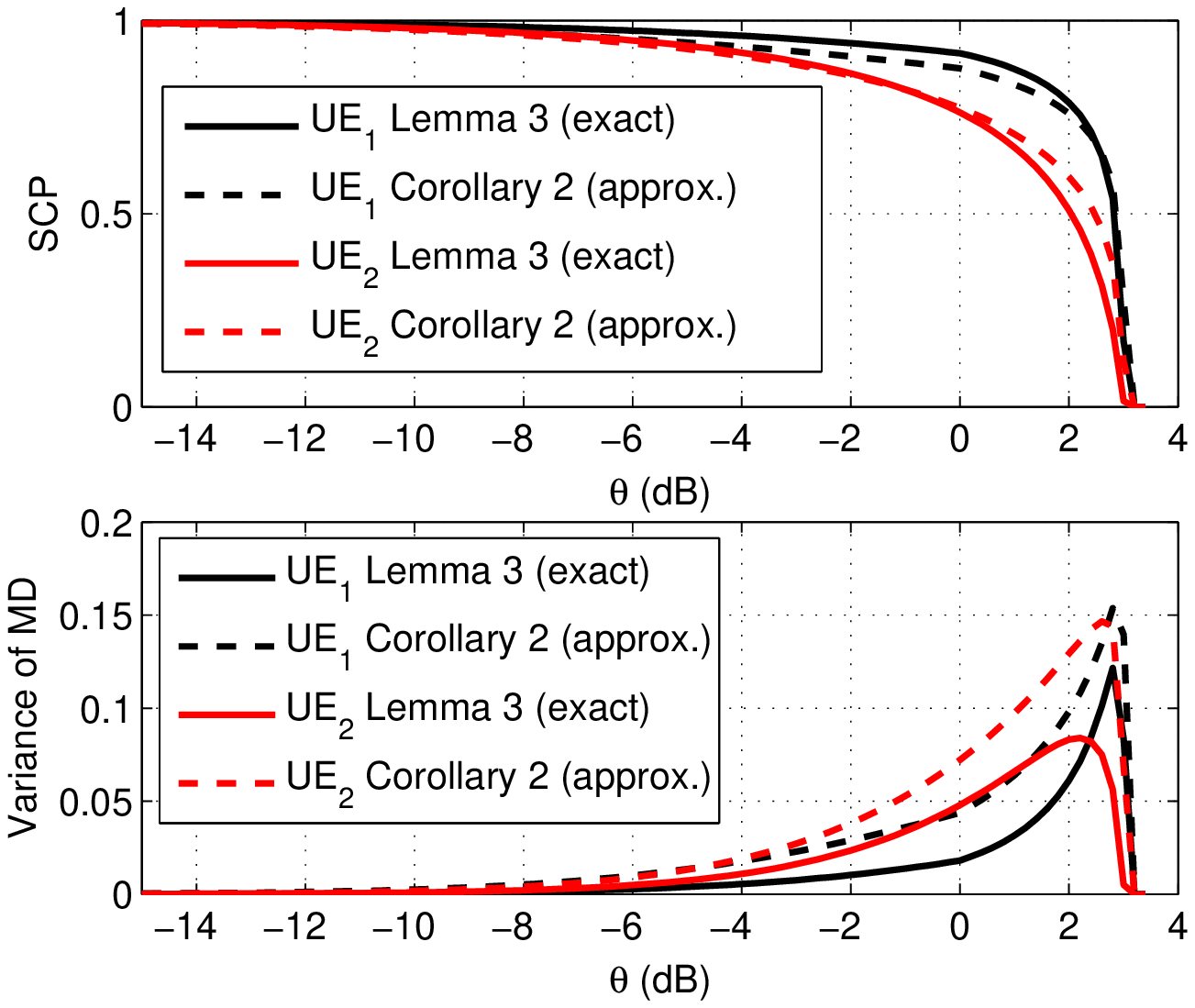}
\caption{SCP and variance of the MD vs. $\theta$ (identical target rate for all UEs) with $P_1=1/3$ for the C-NOMA scheme using the exact and approximate moments of CP.}\label{approx}
\end{minipage}\;\;
\begin{minipage}[t]{0.3333\linewidth}
\centering\includegraphics[width=1\textwidth]{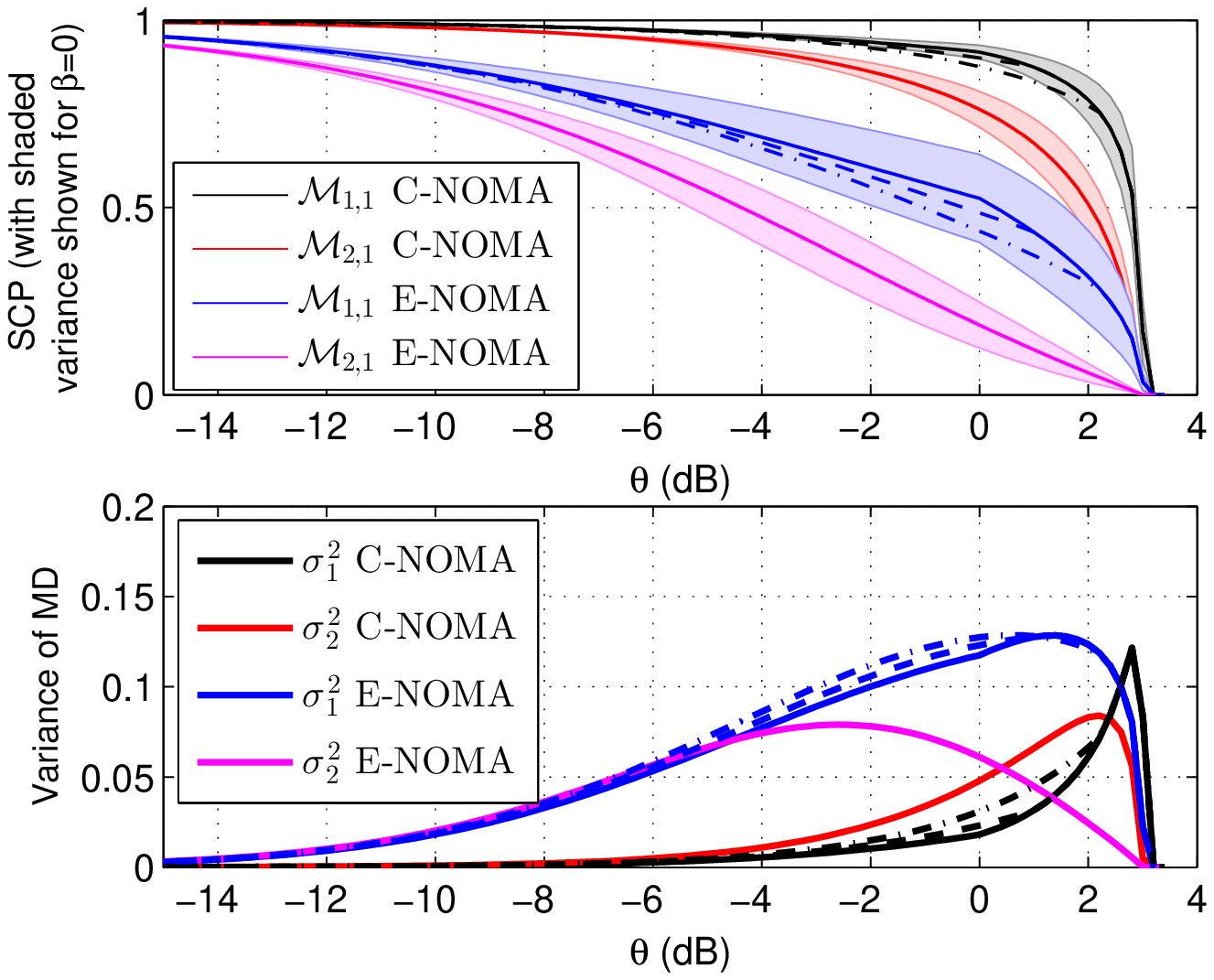}
\caption{SCP and variance of the MD vs. $\theta$ (identical target rate for all UEs) with $P_1=1/3$ for {both} schemes. {Solid lines are for $\beta=0$, dashed for $\beta=0.1$, dash-dot for $\beta=0.2$. Note: the weakest NOMA UE is unaffected by $\beta$.} }\label{covgNvar}
\end{minipage} \;\;
\begin{minipage}[t]{0.33\linewidth}
\centering\includegraphics[width=1\textwidth]{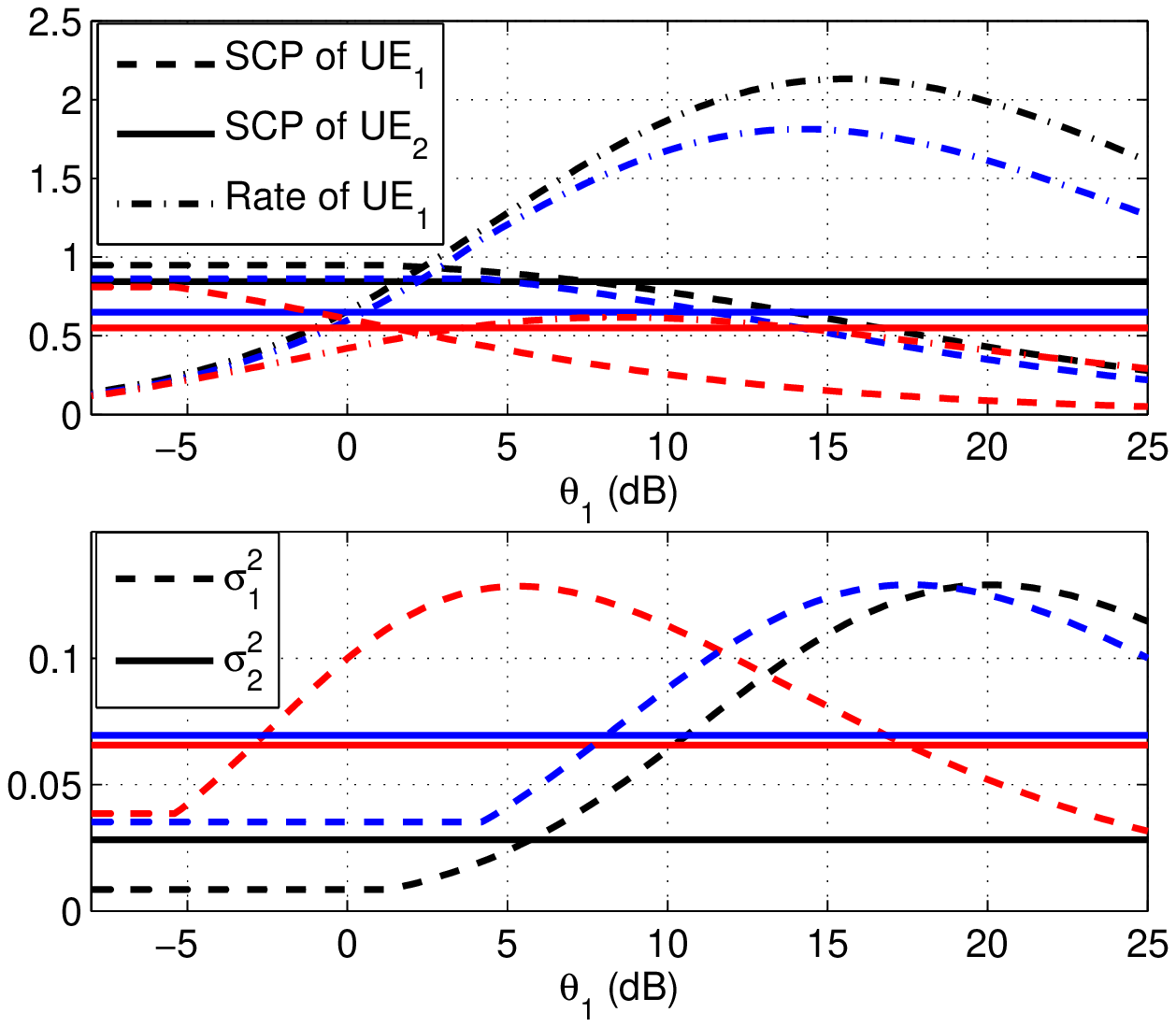}
\caption{SCP and variance of the MD vs. $\theta_1$. For C-NOMA: TMR=0.1 (black) uses $P_2=0.18$ and $\theta_2=-9$ dB, TMR=0.4 (blue) uses $P_2=0.54$ and $\theta_2=-0.7$ dB. For E-NOMA: TMR=0.1 (red) uses $P_2$=0.47 and $\theta_2$=-7 dB.}\label{covgNvar2}
\end{minipage}
\end{figure*}



Fig. \ref{meta} verifies the approximation of the MD in \eqref{meta_beta} using simulations for both schemes with different values of $P_1$. {The approximation is tighter (looser) for C-NOMA (E-NOMA) because of its larger (smaller) interference-exclusion disk with radius $\rho-R_i$ ($R_i$).} {The fraction of $\text{UE}_i$ that attain a given CP is always much larger for C-NOMA when compared to E-NOMA, which highlights the superiority of restricting NOMA to cell-center UEs. When $P_1=0.5$, 98.9\% (92.1\%) of {$\text{UE}_1$ ($\text{UE}_2$)} achieve a CP of at least 0.5 in C-NOMA, while  only 61.5\% (19.9\%) of $\text{UE}_1$ ($\text{UE}_2$) achieve the same CP in E-NOMA.} {Decreasing $P_1$ worsens the performance of $\text{UE}_1$ and improves $\text{UE}_2$; consequently, decreasing $P_1$ in Fig. \ref{meta} increases the fraction of $\text{UE}_2$ that attains a certain CP at the expense of reducing the fraction of $\text{UE}_1$ achieving a given CP}.

%

Fig. \ref{approx} plots the mean and variance of the MD for the NOMA UEs in the C-NOMA scheme. We compare using the moments obtained with and without the approximations \textbf{A1} and \textbf{A2}. We observe that the approximation is tight for the SCP and overestimates the variance, {particularly for $\text{UE}_2$ near the peak}.

Fig. \ref{covgNvar} plots the mean and variance of the MD of the UEs for both schemes using identical RA. We observe that C-NOMA outperforms the E-NOMA scheme in terms of both SCP and variance. {Increasing $\beta$ deteriorates performance of the non-weakest UEs, decreasing SCP and increasing variance. For a given $\beta$,} the higher SCP of the C-NOMA scheme can be attributed to the fact that the UEs are closer to the BS on average than the E-NOMA scheme. {The lower variance is also due to the limited vicinity leading to lower disparity than the E-NOMA model. Furthermore, $\sigma_i^2$ peaks at high $\theta$ for the C-NOMA scheme (corresponding to low SCP); which is not the case for the E-NOMA scheme. This implies the existence of $\theta$ with high SCP and low $\sigma_i^2$ in C-NOMA, thereby highlighting its superiority with careful RA.} {The C-NOMA is also a more consistent scheme as both SCP and variance are better for $\text{UE}_1$ than $\text{UE}_2$; this is not the case for the E-NOMA scheme.}  



Fig. \ref{covgNvar2} plots the mean and variance of the MD for an optimized power and rate adaptation for UE$_2$ such that the total rate is maximized subject to a threshold minimum rate (TMR) constraint. The rate of a UE is defined as the SCP times target rate. RA is done according to the algorithm in \cite{myNOMA_icc} and results in $\text{UE}_2$ having rate equal to the TMR. We also plot the rate of $\text{UE}_1$ in Fig. \ref{covgNvar2}. {In C-NOMA (and E-NOMA, not shown for brevity), increasing the {TMR} increases $\sigma_2^2$ while the peak $\sigma_1^2$ occurs at lower $\theta_1$ but does not change in value.} When the TMR is $0.1$, the SCP of $\text{UE}_2$ and $\sigma_2^2$ are worse for E-NOMA. Although the peak $\sigma_1^2$ is higher for C-NOMA than E-NOMA, at the optimum $\theta_1$ that maximizes the rate of $\text{UE}_1$, $\sigma_1^2$ is lower for C-NOMA. {Other than highlighting the superiority of the C-NOMA scheme, this also emphasizes the importance of optimum RA {for not just the SCP, but also for higher moments of the MD}.}





\section{Conclusion}\label{Conc}
We study the meta distribution of the CCP of NOMA UEs distributed according to two models. Closed form expressions for the moments of the meta distribution in the E-NOMA scheme are derived. The C-NOMA scheme requires a triple integral so we propose approximate moments that reduce to a single integration. Our results show that employing NOMA for cell-center users is significantly more beneficial than using it for all UEs in a cell, thereby motivating the works of \cite{myNOMA_icc,myNOMA_tcom}. We also emphasize the importance of RA in NOMA.

\appendices

\bibliographystyle{IEEEtran}
\bibliography{refsNOMA}

\end{document}